\documentclass[10pt]{article}
\usepackage{amsmath,amssymb}
\usepackage{epsfig}
\usepackage{psfrag}
\begin{document}

\numberwithin{equation}{section}
\newtheorem{theorem}{Theorem}
\newtheorem{proposition}{Proposition}
\newtheorem{remark}{Remark}
\newtheorem{lemma}{Lemma}
\newtheorem{corollary}{Corollary}
\newtheorem{definition}{Definition}
\def\NN{{\mathbb N} }
\def\ZZ{{\mathbb Z} }
\def\RR{{\mathbb R} }
\def\PP{{\mathbb P} }
\def\EE{{\mathbb E} }
\def\dis{\displaystyle}
\def\P{{\mathcal P} }
\def\Q{{\mathcal Q} }

\title{A numerical strategy for coarse-graining two-dimensional
  atomistic models at finite temperature: the membrane case}
\author{X. Blanc$^{1}$ and F. Legoll$^{2,3}$\\
{\footnotesize $^1$ CEA, DAM, DIF,
91297 Arpajon, France}\\
{\footnotesize \tt blanc@ann.jussieu.fr, Xavier.Blanc@cea.fr}\\
{\footnotesize $^2$ Institut Navier, LAMI, \'Ecole Nationale des Ponts et
Chauss\'ees, Universit\'e Paris-Est,}\\
{\footnotesize 6 et 8 avenue Blaise Pascal, 77455 Marne-La-Vall\'ee
Cedex 2, France}\\
{\footnotesize \tt legoll@lami.enpc.fr}\\
{\footnotesize $^3$ INRIA Rocquencourt, MICMAC team-project,}\\
 {\footnotesize Domaine de Voluceau, B.P. 105,
 78153 Le Chesnay Cedex, France}
}
\date{\today}

\maketitle

\begin{abstract}
We present a numerical strategy to compute ensemble averages of
coarse-grained two-dimensional membrane-like models. The approach
consists in generalizing to these two-dimensional models a one-dimensional
strategy exposed in~[Blanc, Le Bris, Legoll, Patz, JNLS
2010,~\cite{bllp}], which is based on applying the ergodic theorem to
Markov chains. This may be considered as a first step towards computing
the constitutive law associated to such models, in the thermodynamic limit.
\end{abstract}

\section{Introduction}
\label{sec:2D}

A standard problem in materials science is the computation of canonical
averages. As an example, consider a system of $P$ particles, at
positions $u = (u^1, \dots, u^P) \in \RR^{3P}$, and assume that its
energy is given by a two-body interaction:
\begin{equation}
  \label{eq:nrj}
E(u) = \frac12 \sum_{i\neq j} W(u^i - u^j).  
\end{equation}
Given an observable $A:\RR^{3P}\to \RR$, and $\beta>0$ the inverse of
the temperature, the canonical ensemble average of $A$ is defined by
(see e.g.~\cite{deak})
\begin{equation}
  \label{eq:average}
  \langle A \rangle = \frac{\dis\int_{\Omega^P} A(u) e^{-\beta E(u)}du
  }{\dis \int_{\Omega^P} e^{-\beta E(u)}du},
\end{equation}
where $\Omega \subset \RR^3$ is the macroscopic domain in which the
particles are assumed to lie. 

Such canonical averages relate the macroscopic properties of a system to
the elementary phenomena at the microscopic scale, which are modelled by
the atomistic energy~\eqref{eq:nrj}. For instance, the bulk pressure in
a fluid is given by~\eqref{eq:average}, for the observable
$$
A(u) = \rho T 
- \frac{1}{3|\Omega|} \sum_{i=1}^P 
u^i \cdot \frac{\partial E}{\partial u^i}(u),
$$
where $T$ is the temperature and $\rho$ the density. If the atoms are
non-interacting, then $E \equiv 0$ and we recover Mariotte's law for ideal
gas. In general, the relation~\eqref{eq:average} (with the above
choice for $A$) allows to compute the pressure of a liquid at a given
density and temperature, and hence to obtain its macroscopic
constitutive law, based on the microscopic model~\eqref{eq:nrj}.  

The integrals in~\eqref{eq:average}
involve $3P$ variables, and this number is extremely large when one
considers a macroscopic sample of matter. Therefore, a direct
computation of $\langle A \rangle$, by standard quadrature rules (using
Gauss integration points), is not 
tractable. Several approaches have been proposed to deal with such
problems (see e.g.~\cite{sampling-review} for a theoretical and
numerical comparison of those). The approach we will follow here is to
use the so-called overdamped Langevin dynamics. This stochastic dynamics
reads
\begin{equation}
  \label{eq:langevin}
  du = - \nabla E (u) dt + \sqrt{2/\beta} \ dB_t,
\end{equation}
where $B_t$ is a Brownian motion in dimension $3P$. This dynamics is
hence a steepest 
descent (according to the forces $- \nabla E (u)$), modified by a random
noise, the magnitude of which is scaled by the temperature: the larger
the temperature, the stronger the noise. The
dynamics~\eqref{eq:langevin} is interesting because of the following
ergodicity result: under general assumptions on the energy $E$,
we have 
\begin{equation}
  \label{eq:langevin-average}
  \langle A \rangle = \lim_{T\to\infty} \frac 1 T \int_0^T A(u(t))dt
\end{equation}
for (almost) all initial conditions $u(0)$. In the long time limit, the
random number $\dis \frac 1 T \int_0^T A(u(t))dt$ converges to a
deterministic number, which is the canonical average of interest.

Thus, in order to compute~\eqref{eq:average}, a standard approach consists in simulating
the evolution of the system according to~\eqref{eq:langevin} (this can
be done even for large systems), and averaging the quantity $A(u(t))$
along the obtained trajectory. The
equation~\eqref{eq:langevin-average} ensures that the time average of
$A(u(t))$ converges to the desired quantity $\langle A \rangle$ in the
long time limit. In practice, the equation~\eqref{eq:langevin} is
numerically integrated with the forward Euler scheme 
$$
u_{m+1} = u_m - \Delta t \, \nabla_u E(u_m) + 
\sqrt{2 \Delta t \beta^{-1}} \ G_m
$$
with the time step $\Delta t$, 
where $G_m$ is a vector of Gaussian random variables in dimension
$3P$ (all components of $G_m$ are independent from each other, and they
are all distributed according to a Gaussian law of mean 0 and variance
1). In turn, following~\eqref{eq:langevin-average}, the canonical 
average~\eqref{eq:average} is approximated by 
$$
\langle A \rangle \approx \lim_{M \to \infty} \frac1M \sum_{m=1}^M A
\left(u_m \right).
$$

We described above how to compute the canonical
average~\eqref{eq:average} using the 
overdamped Langevin dynamics~\eqref{eq:langevin}. We point out that we
give {\em no physical meaning} to this dynamics, which is only
considered here as a {\em numerical tool} to sample the Gibbs measure
and allow efficient computation of canonical averages. 

As pointed out above, several such numerical tools have been proposed in
the literature, and among them, the Langevin equation, which is closer
to physical, Hamiltonian dynamics than the overdamped Langevin
dynamics~\eqref{eq:langevin}. The Langevin dynamics reads
\begin{equation}
\label{eq:full_langevin}
\begin{array}{rcl}
du &=& M^{-1} p \, dt,
\\
dp &=& - \nabla E (u) dt - \gamma M^{-1} p \, dt 
+ \sqrt{2 \gamma /\beta} \ dB_t,
\end{array}
\end{equation}
where $u \in \RR^{3P}$ represents the position of the atoms, $p \in
\RR^{3P}$ their momentum, $M$ is a mass matrix, and $\gamma>0$ is a
parameter. This dynamics 
shares the same ergodicity property as the overdamped Langevin dynamics:
for any $\gamma>0$, the convergence~\eqref{eq:langevin-average} holds
along trajectories of~\eqref{eq:full_langevin}. 

Note that, when
$\gamma=0$, we recover from~\eqref{eq:full_langevin} the standard
Hamiltonian dynamics (Newtonian dynamics). When $\gamma>0$, the Langevin
dynamics includes some friction 
and some random noise, which is, as for the overdamped Langevin
dynamics, scaled by the temperature. Finally, in the limit $\gamma \to
\infty$, and up to some time rescaling, the Langevin
dynamics~\eqref{eq:full_langevin} converges to the overdamped Langevin
dynamics~\eqref{eq:langevin} (see e.g.~\cite[Sec. 2.2.4]{book_fe}).

\bigskip

In many cases, the observable $A$ of interest does not depend on all positions
$u^i$, but only on a few of them. Our aim here is to derive simpler methods
for computing canonical averages in such a case, based on a
coarse-graining strategy. Such methods were used in~\cite{bllp} for
one-dimensional models (see also~\cite{lncse}), and we adapt here the
strategy to a two-dimensional membrane-like model.

One possible approach to address such question is the 
{\em QuasiContinuum Method} (QCM).
It was first introduced in~\cite{qcm5,qcm4}, and further developed
in~\cite{qcm7,miller-tadmor,qcm3,qcm2,qcm6,qcm1}, in the zero
temperature case. It has been theoretically studied
in~\cite{anitescu,arndt-griebel,arndt-luskin,qcm-m2an,qcm-sinica,archive,dobson,e2,e4,e3,lin1,lin2,ortner}.
See also~\cite{m2an-review,berlin} for some review articles. The method
was adapted to the finite temperature case in~\cite{hot-qc}, resulting
in a coarse-graining strategy to compute ensemble averages. Similar
strategies were also proposed in~\cite{ceder,LeSar}.

In essence, any
coarse-graining approach aims at using the fact that $A$ depends only on
the positions of some representative atoms at positions $u_r \in
\Omega^{P_r}$, with $P_r \ll P$. This is usually done by writing formally
$$
\langle A \rangle 
=
\frac{1}{Z_P} \int_{\Omega^{P}} A(u_r) e^{-\beta E(u)} du  
=
\frac{1}{Z_{P_r}} \int_{\Omega^{P_r}} A(u_r) e^{-\beta E_{CG}(u_r)} du_r,
$$
where
$$
Z_P = \int_{\Omega^P} \exp\left(-\beta E(u)\right) du, \quad Z_{P_r} =
\int_{\Omega^{P_r}} \exp\left(-\beta E_{CG}(u_r)\right) du_r,$$
and where
\begin{equation}
\label{eq:cg-energy}
E_{CG}(u_r) = -\frac 1 \beta \ln \left[\int_{\Omega^{P-P_r}}
  \exp\left(-\beta E(u) \right) d\hat u_r\right]
\end{equation}
is the coarse-grained energy. In~\eqref{eq:cg-energy}, the notation
$d\hat u_r$ indicates that we integrate over all variables except $u_r$.
The question then reduces to computing the coarse-grained energy
$E_{CG}$, which can be considered as the free energy of the system, for
the collective variable (or reaction coordinate) $u_r$.

Note that $E_{CG}(u_r)$ depends on $P$, and is in general difficult to 
compute for a given value of $P_r$ and $P$. However, one may hope that
its expression simplifies in the limit when
$$
P_r \text{ is fixed, } P \to \infty.
$$
This is the case for instance in dimension one, in the case of
nearest-neighbour interactions, as it is recalled in Section~\ref{ssec:1d}
below. 

Rather than fixing $\beta$ and letting $P \to \infty$, as described
above, another
asymptotic regime is to keep $P$ and $P_r$ fixed and let $\beta$ go to
$+\infty$ (vanishing temperature). This is the regime considered by the 
QCM approach, which is based, roughly speaking, on using a harmonic
approximation of $E(u)$ in~\eqref{eq:cg-energy}. This then allows
to compute explicitly $E_{CG}(u_r)$.

\medskip

The sequel of this article is organized as follows. In
Section~\ref{ssec:1d}, we consider one-dimensional chains of atoms, and
recall our results of~\cite{bllp}. We next present in
Section~\ref{ssec:setting-2d} the two-dimensional membrane-like models
we consider here, and the questions we are after. Section~\ref{sec:theo}
is dedicated to the derivation of our numerical strategy, which is very
much inspired by our previous works on one-dimensional models. The
outcome of this derivation is an algorithm, that we have summarized in
Section~\ref{sec:algo}. Numerical results obtained with this algorithm
are presented (and compared with the reference model results) in
Section~\ref{sec:num}. 

\subsection{Dimension one}
\label{ssec:1d}

In~\cite{bllp}, we proposed a methodology to deal with
one-dimensional models. Although the method allows to handle models with any
finite range interaction, we recall it here in the special case of
nearest-neighbour and next-to-nearest-neighbour interaction, refering
to~\cite{bllp,lncse} for more details. 

Consider a chain of $P$ atoms at positions $u^i \in
\RR$, with an energy of nearest-neighbour type:
\begin{equation}
\label{eq:energy-1d-NN}
E(u) = \sum_{i=0}^{P-1} W\left(\frac{u^{i+1} - u^i}{h}\right),
\end{equation}
where $h=1/P$ is the average spacing of the atoms, and is assumed here
to be equal to the characteristic length of the potential (as the equilibrium
length of $z \mapsto W(z)$ is assumed to be of order 1, the equilibrium
length of $z \mapsto W(z/h)$ is of the order of $h$). To remove the
translation invariance of the energy, we set $u^0 = 0$. 

As explained above, we consider the canonical average of observables
depending only on a few atoms of the system. To simplify the setting, we
assume that $A$ depends only on $u^P$. Thus, the quantity we wish to
compute is
$$
\langle A \rangle_P 
= 
\frac{1}{Z_P} \int_{\RR^P} A\left(u^P\right) \exp\left[
-\beta \sum_{i=0}^{P-1} W\left(\frac{u^{i+1} - u^i}{h}\right)
\right] du,
$$
where
$\dis
Z_P = \int_{\RR^P} \exp\left[
-\beta \sum_{i=0}^{P-1} W\left(\frac{u^{i+1} - u^i}{h}\right)
\right] du$.
More precisely, we are going to compute the limit of $\langle A
\rangle_P$ when $P \to \infty$, that is in the thermodynamic limit, when
the number of atoms present in the system diverges. 

Changing variables according to
\begin{equation}
\label{eq:change}
y_i := \frac{u^{i} - u^{i-1}}{h}, \ \ i=1, \ldots, P,
\end{equation}
which implies $\dis u^P = \frac{1}{P} \sum_{i=1}^P y_i$, we have
\begin{equation}
\label{eq:average-1d}
\langle A \rangle_P = \frac{1}{Z_P} \int_{\RR^P} A\left(\frac{1}{P}
  \sum_{i=1}^P y_i\right) \exp\left[ -\beta \sum_{i=1}^P W(y_i) \right]
dy,
\end{equation}
with
$$
Z_P =  \int_{\RR^P} \exp\left[ -\beta\sum_{i=1}^P W(y_i) \right] dy.
$$
Thus,~\eqref{eq:average-1d} may be interpreted as the expectation value
of a sequence of independent identically distributed (i.i.d.) variables:
$$
\langle A \rangle_P 
= 
\EE \left[A \left(\frac{1}{P} \sum_{i=1}^P Y_i \right) \right], 
$$
where $Y_i$ is a sequence of i.i.d. variables distributed according to
the law $d\mu(y) = Z^{-1} \exp(-\beta W(y)) dy$, and $Z = \int_\RR
\exp(-\beta W(y))dy$. For finite (and large) $P$, the quantity $\langle
A \rangle_P$ is not particularly easy to compute. However, its limit
when $P \to \infty$ is easy to compute. Indeed, applying the law of
large numbers~\cite{shiryaev}, we know that 
$\dis \frac{1}{P} \sum_{i=1}^P Y_i$ converges to the mean $\EE(Y_i)$ of
the random variables $Y_i$, and we hence have (see~\cite{bllp}):
\begin{equation}
\label{eq:lln_nn}
\lim_{P\to \infty} \langle A \rangle_P = A(y^\star), 
\quad 
y^\star := \EE(Y_i) = \frac{\dis
\int_\RR y \exp(-\beta W(y)) dy}{\dis \int_\RR  \exp(-\beta W(y)) dy}.
\end{equation}

\begin{remark}
Choosing $A(y) = y$ in the above relation, we observe that $y^\star$
satisfies $\dis y^\star = \lim_{P\to \infty} \langle u^P \rangle_P$. 
In the specific one-dimensional nearest-neighbour case discussed here, we actually have 
$y^\star = \langle u^P \rangle_P$ for any $P$. Likewise, for any $P$ and
any $1 \leq i \leq P$, we have $\dis y^\star = \langle y_i \rangle_P =
\langle \frac{u^i - u^{i-1}}{h} \rangle_P$.
\end{remark}

The property~\eqref{eq:lln_nn} holds under mild assumptions on $A$ and
$W$. The only important constraint is that $\exp(-\beta W)$ should be
integrable over the real line. Actually, finer results may be easily
obtained, as for instance Theorem~1 of~\cite{bllp}.

\bigskip

In the case of next-to-nearest-neighbour interaction, it is possible to
apply the same kind of strategy, but it is much more involved. Assume
now that, instead of~\eqref{eq:energy-1d-NN}, the energy reads
\begin{equation}
  \label{eq:energy-1d-NNN}
  E(u) = \sum_{i=0}^{P-1} W_1\left(\frac{u^{i+1}- u^i}h\right) + \sum_{i=1}^{P-1}
  W_2\left(\frac{u^{i+1} - u^{i-1}}h \right).
\end{equation}
Again, we assume that the observable $A$ depends only on the right-end
atom position $u^P$, and that $u^0=0$.  
Then, using the same change of variables~\eqref{eq:change}, we have
\begin{eqnarray}
\nonumber
\langle A \rangle_P
&=&
\frac{1}{Z_P} \int_{\RR^P} A\left(u^P\right) \exp\left[
-\beta E(u) \right] du
\\
\nonumber
&=&
\frac{1}{Z_P} \int_{\RR^P} A\left(\frac{1}{P} \sum_{i=1}^P y_i\right) 
\exp\left[-\beta \sum_{i=1}^{P} W_1(y_i) \right] 
\\
&& \hspace{3cm} \times \exp\left[ -
    \beta \sum_{i=1}^{P-1}W_2(y_i + y_{i+1}) \right] dy.
\label{eq:average-NNN}
\end{eqnarray}
Here, instead of i.i.d. variables, we may identify a Markov chain
structure. Introduce indeed
$$
k(a,b) = \exp \left[-\frac{\beta}{2} W_1(a) - \frac{\beta}{2} W_1(b) -\beta
    W_2(a+b) \right].
$$
Then we can recast~\eqref{eq:average-NNN} as
\begin{multline}
\langle A \rangle_P
=
\frac{1}{Z_P} \int_{\RR^P} A\left(\frac{1}{P} \sum_{i=1}^P y_i\right) 
\exp\left[-\frac{\beta}{2} W_1(y_1) -\frac{\beta}{2} W_1(y_P) \right] 
\\
\times \prod_{i=1}^{P-1} k(y_i,y_{i+1}) dy.
\label{eq:average-NNN-bis}
\end{multline}
For the sake of simplicity, assume now that $\dis \int_\RR k(a,b) \, db =
1$. Then we can introduce a Markov chain $\left\{ Y_i \right\}_{i \geq
  0}$ of kernel $k$, namely such that
$k(a,b) = \PP(Y_{i+1} = b \ | \ Y_i = a)$, and
recast~\eqref{eq:average-NNN-bis} as
$$
\langle A \rangle_P
=
\frac{\EE \left[ A\left(\frac{1}{P} \sum_{i=1}^P Y_i\right) 
\exp\left(-\frac{\beta}{2} W_1(Y_1) -\frac{\beta}{2} W_1(Y_P) \right)
\right]}
{\EE \left[  
\exp\left(-\frac{\beta}{2} W_1(Y_1) -\frac{\beta}{2} W_1(Y_P) \right) 
\right]}.
$$
Applying next the law of large numbers (this time for Markov chains,
rather than for i.i.d. variables as previously), we can identify
$\dis \lim_{P \to \infty} \langle A \rangle_P$. 

In general, the quantity $\dis \int_\RR k(a,b) \, db$ depends on $a$,
and is not
equal to 1. So the above discussion needs to be amended. It turns out that it
is nevertheless possible to identify a Markov chain structure
in~\eqref{eq:average-NNN}. Hence, applying asymptotic theorems (namely
law of large numbers type results) for this setting (see
e.g.~\cite{shiryaev}), we have (see~\cite{bllp} for the details): 
\begin{equation}
  \label{eq:lin-NNN}
  \lim_{P\to \infty} \langle A \rangle_P = A(y^\star), 
\quad 
y^\star = \int_\RR y \left( \psi_1(y) \right)^2 dy,
\end{equation}
where $\psi_1^2$ is the invariant measure of the transition kernel of
the Markov chain, that is, the solution of
\begin{equation}
  \label{eq:psi1-NNN}
  \P \psi_1 = \lambda \psi_1, \quad \lambda = \max \text{Spectrum}(\P),
\end{equation}
where the transition operator $\P$ is defined on functions $\phi : \RR
\to \RR$ by
$$
\left(\P\phi\right)(x) = \int_\RR
  \phi(y) \exp\left[-\frac\beta 2 W_1(x) - \frac\beta 2 W_1(y)-\beta
    W_2(x+y) \right]dy.
$$
Note that, using standard tools of spectral theory, it is possible 
to prove that, under some reasonable assumptions,
problem~\eqref{eq:psi1-NNN} has a unique solution (we refer
to~\cite{bllp} for the details).

\begin{remark}
\label{rq:y*-NNN}
As in the nearest-neighbour case, choosing $A(y) = y$
in~\eqref{eq:lin-NNN}, we have $\dis y^\star =
\lim_{P\to\infty} \langle u^P \rangle_P$. Note that we also have
$$
y^\star = \lim_{P\to\infty} \langle y_i \rangle_P
= \lim_{P\to\infty} \langle \frac{u^i - u^{i-1}}{h} \rangle_P
$$
for any $1 \ll i \ll P$. Otherwise stated, $y^\star$ is the average
(rescaled) elongation of any bond which is in the bulk of the chain.
\end{remark}

Before moving on to the two-dimensional setting, let us briefly describe
one application of the above strategy. We have considered until now
chains of atoms at rest, that is, submitted to {\em no external force}. The
energy of the system reads~\eqref{eq:energy-1d-NN}
(or~\eqref{eq:energy-1d-NNN}). Choosing $A(u) = u$ in~\eqref{eq:lln_nn}
(or~\eqref{eq:lin-NNN}), we see that $y^\star$ is the average elongation of
the system, at a given finite temperature (macroscopic equilibrium
length). It turns out that the above 
strategy can be easily extended to treat the case when some external
force is applied to the right-end atom of the chain (at position $u^P$),
while we again impose the position of the left-end atom: $u^0=0$. In this case,
the energy, rather than~\eqref{eq:energy-1d-NN}, reads
$$
E_f(u) = 
\sum_{i=0}^{P-1} W\left(\frac{u^{i+1} - u^i}{h}\right) - f \frac{u^P}{h},
$$
and the average length of the chain, at the inverse temperature $\beta$,
and when a force $f$ is applied at the right-end, reads
$$
\langle A \rangle^f_P 
= 
\frac{1}{Z_P} \int_{\RR^P} A\left(u^P\right) \exp\left[
-\beta E_f(u) \right] du,
$$
for $A(u) = u$. Following the same arguments as above (see~\cite{lncse}
for details), we can compute 
$\dis y^\star_f = \lim_{P \to \infty} \langle u^P \rangle^f_P$, which is
the average {\em elongation} of the system when submitted to the
external {\em force} $f$. We thus have computed the constitutive law of
the chain (namely the relationship between force and elongation), at
finite temperature.

\subsection{Setting of the problem}
\label{ssec:setting-2d}

We now turn to describing the problem we address in this article. The
above treatment of the one-dimensional setting will be used as a 
bottom-line to treat the case at hand here, namely two-dimensional
membrane-like models. In contrast to the one-dimensional case,
we are not able to prove the convergences we claim. However, we believe
that the one-dimensional proofs are sufficient to allow for the
approximations we are going to make. Extensive numerical results
reported in Section~\ref{sec:num} below show the accuracy of our
approximations. 

\medskip

\begin{figure}[h]
\centerline{
\input{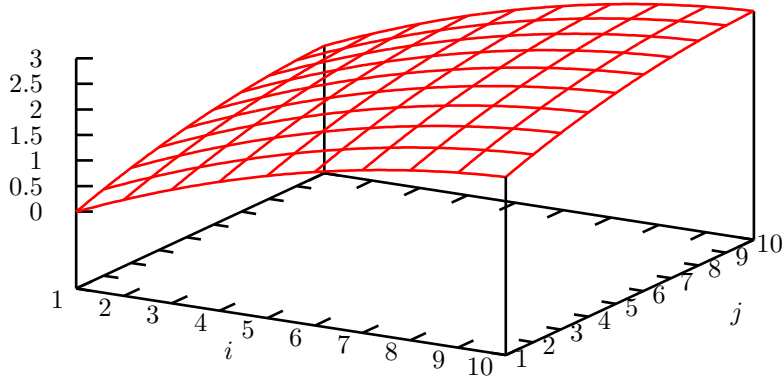}
}
\caption{Schematic representation of the system: atoms are set on a
  two-dimensional lattice, and have a unique degree of freedom, which is
  their height.} 
\label{fig:nappe}
\end{figure}

We consider a two-dimensional atomistic system, where each atom has a
{\em scalar} degree of freedom. The mechanical interpretation is that
the system is a {\em membrane}, and the degree of freedom is the height
of the atom (see Figure~\ref{fig:nappe}).  

We hence consider a two-dimensional set of $P = (N+1)^2$ atoms at
positions 
$$
u^{i,j} \in \RR, \quad 0\leq i,j\leq N.
$$
We assume a nearest-neighbour interaction, and thus write the energy as
\begin{equation}
\label{eq:nrj2d}
E \left( \left\{ u^{i,j} \right\} \right) = \sum_{0\leq i,j \leq N}
\sum_{\begin{array}{c} 
\scriptstyle {0\leq k,l\leq N} 
\\ 
\scriptstyle {i \leq k \leq i+1, \ j \leq l \leq j+1 }
\\
\scriptstyle {(k-i) + (l-j) =1 } 
\end{array} } W\left(\frac{u^{k,l} - u^{i,j}}h\right).
\end{equation}
Note that, as in~\eqref{eq:energy-1d-NN}, we have rescaled the relative
position by the factor $h=1/N$. Likewise, we consider in the above sum
oriented bonds, from left to right or bottom to top (but not the
contrary). Finally, as in the one-dimensional case, we do not have to
assume $W$ to be an even function. The energy~\eqref{eq:nrj2d} 
is translation invariant, so we fix in the sequel $u^{0,0} = 0$ to
remove this invariance. Our aim
is to compute the canonical average of an observable $A$ that only
depends on $u^{N,N}$, with respect to the Gibbs measure associated to
the energy~\eqref{eq:nrj2d}. This quantity reads
\begin{equation}
\label{eq:aim}
\langle A \rangle_N =
Z_N^{-1} \int_{\RR^{(N+1)^2-1}} A\left(u^{N,N}\right) \ 
\overline{\mu}_N \left(\left\{u^{i,j}\right\}\right) \,
\prod_{0\leq i,j \leq N, \ (i,j) \neq (0,0)} du^{i,j},
\end{equation}
where 
\begin{equation}
\label{eq:gibbs}
\overline{\mu}_N\left(\left\{u^{i,j}\right\}\right) = \exp(-\beta 
E \left( \left\{ u^{i,j} \right\} \right) )
\end{equation}
is the non-normalized Gibbs measure, and 
$$
Z_N = \int_{\RR^{(N+1)^2-1}} \overline{\mu}_N \left(\left\{u^{i,j}\right\}\right) \,
\prod_{0\leq i,j \leq N, \ (i,j) \neq (0,0)} du^{i,j}
$$
is the normalization constant. Our exact aim is to compute $\langle A
\rangle_N$ in the {\em thermodynamics limit}, namely $\dis \lim_{N \to
  \infty} \langle A \rangle_N$.

\medskip

As in~\cite{bllp}, we aim at describing a numerical strategy to compute 
$\dis \lim_{N \to \infty} \langle A \rangle_N$ that builds on the
specificity that the observable $A$ under study depends {\em only} on
$u^{N,N}$, and not on {\em all} the atom positions 
$\left\{ u^{i,j} \right\}_{0 \leq i,j \leq N}$. As we will see below, we
are also able to handle observables that depend only of $u^{0,N}$ or
$u^{N,0}$, and, more generally, of $u^{0,N}$, $u^{N,0}$ and $u^{N,N}$.
We build this strategy in Section~\ref{sec:theo}, and obtain the
algorithm summarized in Section~\ref{sec:algo}. Numerical results
are reported in Section~\ref{sec:num}.

\begin{remark}
In dimension one, in addition to computing
$\dis \lim_{P \to \infty} \langle A \rangle_P$ (as recalled in
Section~\ref{ssec:1d}), it is also possible, using
a large deviation principle strategy, to compute the coarse-grained energy
$E_{CG}$ defined by~\eqref{eq:cg-energy}. We refer to~\cite{bllp,lncse}
for the details. In dimension two, such a problem is known to be very
difficult, and very few theoretical results are known. See for
instance~\cite{funaki-spohn,dgi_00,gos_01,ns_97}. This is why we
concentrate here on the simpler question of computing the average of an
observable. 
\end{remark}

\section{Numerical strategy}
\label{sec:theo}

This section is dedicated to the construction of our numerical
strategy. It is quite involved from the technical viewpoint. For the
sake of clarity, we have summarized in Section~\ref{sec:algo} below the
resulting algorithm which has eventually to be implemented.

\smallskip

We first rewrite the energy~\eqref{eq:nrj2d} using the increments
between the atoms. Introducing 
\begin{eqnarray*}
x_{i,j} &=& \frac{u^{i,j} - u^{i-1,j}}h, \quad 1 \leq i \leq N, \quad 
0 \leq j \leq N,
\\
y_{i,j} &=&
\frac{u^{i,j} - u^{i,j-1}}h, \quad 0 \leq i \leq N, \quad 
1 \leq j \leq N,
\end{eqnarray*}
we have
\begin{equation}
\label{eq:nrj2d2}
E\left(\left\{u^{i,j}\right\}\right) 
=
\sum_{1\leq i,j \leq N} W(x_{i,j}) + W(y_{i,j}) 
+ \sum_{i=1}^{N} W(x_{i,0}) + \sum_{j=1}^{N} W(y_{0,j}).
\end{equation}
This change of variables is not one-to-one: $(N+1)^2-1$ variables are involved
in~\eqref{eq:nrj2d}, whereas $2N^2 + 2N$ variables are involved
in~\eqref{eq:nrj2d2}. The difference is accounted for by imposing the
geometric constraints 
$$
x_{i,j} + y_{i-1,j} = y_{i,j} + x_{i,j-1},
$$
for $1\leq i,j \leq N$. 

In the variables $(x_{i,j},y_{i,j})$, the non-normalized Gibbs measure
$\overline{\mu}_N$ reads
\begin{multline} 
\label{eq:gibbs2d}
\prod_{1\leq i,j\leq N} f(x_{i,j}) \, f(y_{i,j}) \ \delta_0\left(x_{i,j} +
y_{i-1,j} - y_{i,j} - x_{i,j-1} \right) 
\\
\times \prod_{1\leq i \leq N} f(x_{i,0}) \prod_{1\leq j
\leq N} f(y_{0,j}),
\end{multline}
with 
$$
f(x) = \exp(-\beta W(x)),
$$ 
and we recast~\eqref{eq:aim} as
\begin{multline}
\label{eq:aim2}
\langle A \rangle_N
=
Z_N^{-1} \int A\left(\frac{1}{N} \sum_{i=1}^N (x_{i,i}+y_{i,i}) \right) 
\prod_{1\leq i \leq N} f(x_{i,0}) \prod_{1\leq j
\leq N} f(y_{0,j})
\\
\times
\prod_{1\leq i,j\leq N} f(x_{i,j}) \, f(y_{i,j}) \ \delta_0\left(x_{i,j} +
y_{i-1,j} - y_{i,j} - x_{i,j-1} \right). 
\end{multline}
Based on our one-dimensional study (recalled in Section~\ref{ssec:1d}),
our first assumption is 
\begin{equation}
\label{eq:H1}
\begin{array}{c}
\text{\textbf{[H1]}} \quad \text{the boundary terms of the Gibbs measure}
\\
\text{(second line of~\eqref{eq:gibbs2d}) are disregarded.}
\end{array}
\end{equation}
Thus, we approximate~\eqref{eq:aim2} by
\begin{equation}
\label{eq:hyp1}
\langle A \rangle_N
\approx
\langle A \rangle_N^{\rm No Boundr},
\end{equation}
where
$$
\langle A \rangle_N^{\rm No Boundr} 
=
Z_N^{-1} \int A\left(\frac{1}{N}
 \sum_{i=1}^N (x_{i,i}+y_{i,i}) \right) 
\mu_N \left(\left\{x_{i,j}\right\}, \left\{y_{i,j}\right\} \right),
$$
with
$$
\mu_N \left(\left\{x_{i,j}\right\}, \left\{y_{i,j}\right\} \right) =
\!\!\! \prod_{1\leq i,j\leq N}\!\!\! f(x_{i,j}) f(y_{i,j}) \delta_0\left(x_{i,j} +
y_{i-1,j} - y_{i,j} - x_{i,j-1} \right). 
$$
Again based on the one-dimensional study, we assume that we can
apply~\eqref{eq:lin-NNN} and Remark~\ref{rq:y*-NNN}, so that 
\begin{equation}
\label{eq:limerg}
\text{\textbf{[H2]}} \quad 
\lim_{N\to +\infty} \langle A \rangle_N^{\rm No Boundr} = A(x^\star + y^\star),
\end{equation}
where
\begin{align}\label{eq:x*2d}
x^\star = \lim_{N\to +\infty} Z_N^{-1} \int x_{k,l} \
\mu_N \left(\left\{x_{i,j}\right\}, \left\{y_{i,j}\right\} \right),
\\
\label{eq:y*2d}
y^\star = \lim_{N\to +\infty} Z_N^{-1} \int y_{k,l} \
\mu_N \left(\left\{x_{i,j}\right\}, \left\{y_{i,j}\right\} \right),
\end{align}
where $k$ and $l$ are any integers such that $1\ll k,l \ll N$. We
implicitly assume in~\eqref{eq:x*2d}-\eqref{eq:y*2d} that $x^\star$ and
$y^\star$ do not depend on $k,l$, provided $1\ll k,l \ll
N$. Collecting~\eqref{eq:hyp1} and~\eqref{eq:limerg}, we thus write
\begin{equation}
\label{eq:motivation}
\lim_{N\to +\infty} \langle A \rangle_N \approx A(x^\star + y^\star).
\end{equation}
All the sequel of this section is dedicated to the computation of (an
approximation of) $x^\star$ and $y^\star$.

\begin{remark}
\label{rem:coins}
Based on the knowledge of $x^\star$ and $y^\star$, we can compute $A(x^\star +
y^\star)$, which is assumed to be an approximation of the average of
$A(u^{N,N})$, according to~\eqref{eq:motivation}. Using the same
assumptions, our strategy leads us to approximate the average of $A(u^{N,0})$
and $A(u^{0,N})$ by $A(x^\star)$ and $A(y^\star)$, respectively. We will
show in Section~\ref{sec:num} that $A(x^\star + y^\star)$, $A(x^\star)$
and $A(y^\star)$ are indeed good approximations of the average of 
$A(u^{N,N})$, $A(u^{N,0})$ and $A(u^{0,N})$, respectively.

As pointed out above, our strategy can straighforwardly be extended to
the case when the observable $A$ depends on $u^{N,N}$, $u^{N,0}$ {\em and}
$u^{0,N}$, although we have not numerically tested such cases in the
sequel. 
\end{remark}

\subsection{Numerical computation of $x^\star$ and $y^\star$}

Our strategy consists in setting an appropriate Markov chain
structure, as we did for the one-dimensional next-to-nearest-neighbour
interaction case (see Section~\ref{ssec:1d}). The variables $x_{i,j}$
in~\eqref{eq:x*2d} are indexed by 
$(i,j)$, which belongs to a subset of $\ZZ^2$, on which there is no
natural ordering. Rather than considering the variables $x_{i,j}$, we
are going to consider the row of variables 
$\left\{ x_{i,j} \right\}_{1 \leq i \leq N}$, indexed by $j$. The
advantage is that there is a natural ordering between these rows of
variables. 

We now proceed in details. For any $1 \leq j \leq N$, define
$$
x_{:,j} = (x_{1,j}, \dots , x_{N,j}) \in \RR^N, 
\quad
y_{:,j} = (y_{1,j}, \dots , y_{N,j}) \in \RR^N.
$$
We thus have 
\begin{eqnarray}
\label{eq:muN2d}
\mu_N &=& \prod_{j=1}^N F(x_{:,j}) \, F(y_{:,j}) \ \delta_0(x_{:,j}
+ \tau y_{:,j} - y_{:,j} - x_{:,j-1}),
\end{eqnarray}
where $\tau$ is the Bernoulli shift, i.e. $(\tau y)_i = y_{i-1},$ and $F$
is the direct product 
$$
F(x_{:,j}) = \prod_{i=1}^N f(x_{i,j}), \quad 
F(y_{:,j}) = \prod_{i=1}^N f(y_{i,j}).
$$
With this notation,~\eqref{eq:x*2d} reads
\begin{equation}
\label{eq:youpi}
x^\star = \lim_{N\to +\infty} Z_N^{-1} \int x_{k,l} \
\prod_{j=1}^N F(x_{:,j}) \, F(y_{:,j}) \ \delta_0(x_{:,j}
+ \tau y_{:,j} - y_{:,j} - x_{:,j-1}).
\end{equation}
We now follow a strategy based on the so-called transfer operators (or
transfer matrices; see for instance~\cite[Chapter I, Section 12]{shiryaev}).
Let us define the transfer operator $\P$ by 
\begin{eqnarray*}
(\P\varphi)(x,y) 
&=& 
F(x) F(y) \int_{\RR^N \times \RR^N} \varphi(t,z) \ 
\delta_0(t+\tau z - z -x) \ dz \ dt
\\
&=&
F(x) F(y) \int_{\RR^N} \varphi(x+z-\tau z,z) \ dz
\end{eqnarray*}
for any function $\varphi$ defined on $\RR^N \times \RR^N$. The adjoint
operator reads 
$$
(\P^\star\varphi)(x,y) = F(x+\tau y - y) \int_{\RR^N} F(z) \ 
\varphi(x+\tau y - y, z) \ dz.
$$
We can compute the law of $(x_{:,l} , y_{:,l})$ by
integrating~\eqref{eq:muN2d} over all variables except $(x_{:,l} ,
y_{:,l})$. The interest of introducing $\P$ and $\P^\star$ is that we
can now recast~\eqref{eq:youpi} in the simple form 
\begin{equation}
\label{eq:youpi2}
x^\star = \lim_{N\to +\infty} \frac{\dis
\int_{\RR^N \times \RR^N} x_{k,l} \ \mu_l(x_{:,l}, y_{:,l}) \ dx_{:,l} \ dy_{:,l}
}{\dis
\int_{\RR^N \times \RR^N} \mu_l(x_{:,l}, y_{:,l}) \ dx_{:,l} \ dy_{:,l}
}
\end{equation}
with
\begin{equation}
\label{eq:muj}
\mu_l(x_{:,l}, y_{:,l}) = \left(\P^{N-l} \varphi_0\right)(x_{:,l},
y_{:,l}) \ \left((\P^\star)^{l-1} \varphi_1\right)(x_{:,l}, y_{:,l}),
\end{equation}
where $\varphi_0(x_{:,N}, y_{:,N}) = F(x_{:,N}) \, F(y_{:,N})$ and
$\varphi_1 = 1$. Assuming $N$ large and
$1\ll l \ll N$, the iterations of $\P$ and $\P^\star$ converge, up to
renormalization, to the spectral projection on the highest eigenvector
of $\P$ and $\P^\star$, respectively. Hence, the law of $(x_{:,l} ,
y_{:,l})$ is 
\begin{equation}
\label{eq:argh}
\mu_l(x_{:,l}, y_{:,l}) = \frac{
\mu(x_{:,l}, y_{:,l}) \ \mu^\star(x_{:,l}, y_{:,l})
}{
\int_{\RR^N \times \RR^N} \mu(x_{:,l}, y_{:,l}) \ 
\mu^\star(x_{:,l}, y_{:,l}) \ dx_{:,l} \ dy_{:,l}
}
\end{equation}
where $\mu$ and $\mu^\star$ are functions defined on $\RR^N \times
\RR^N$ satisfying 
\begin{equation}
\label{eq:vp2d}
\left\{ \begin{array}{l}
\P\mu = \lambda \mu, \\ 
\P^\star \mu^\star = \lambda \mu^\star,
\end{array}\right.\quad \lambda = \sup
\text{Spectrum} (\P) = \sup
\text{Spectrum} (\P^\star). 
\end{equation}

\begin{remark}
The operators $\P$ and $\P^\star$ depend on $N$, since they act on
functions of $2N$ variables. Hence, the highest eigenvalue $\lambda$ as
well as the associated eigenvectors $\mu$ and $\mu^\star$ depend on $N$. To
simplify the notation, we have not made this dependence explicit
in~\eqref{eq:vp2d}. 
\end{remark}

Assuming that the highest eigenvalue of $\P$ is simple and isolated, the
equation~\eqref{eq:youpi2} yields 
\begin{eqnarray}
\nonumber
x^\star &=& \lim_{N\to +\infty} \frac{\dis
\int_{\RR^N \times \RR^N} x_{k,l} \ \mu(x_{:,l}, y_{:,l}) \ 
\mu^\star(x_{:,l}, y_{:,l}) \ dx_{:,l} \ dy_{:,l}
}{\dis
\int_{\RR^N \times \RR^N} \mu(x_{:,l}, y_{:,l}) \ \mu^\star(x_{:,l},
y_{:,l}) \ dx_{:,l} \ dy_{:,l}
}
\\
&=&
\label{eq:youpi3}
\lim_{N\to +\infty} \frac{\dis
\int_{\RR^N \times \RR^N} x_k \ \mu(x,y) \ \mu^\star(x,y) \ dx \ dy
}{\dis
\int_{\RR^N \times \RR^N} \mu(x,y) \ \mu^\star(x,y) \ dx \ dy
}.
\end{eqnarray}
In the above integrals, $x$ denotes the vector $(x_1,\ldots,x_N) \in
\RR^N$, and $1 \ll k \ll N$. 

\begin{remark}
Note that approximating~\eqref{eq:youpi2}-\eqref{eq:muj}
by~\eqref{eq:youpi3} consists in passing to the limit $N \to \infty$ faster
in the vertical direction than in the horizontal direction. Indeed, the
operator $\P$
acts on functions of $2N$ variables, which correspond to increments
$\left\{ x_{i,j}, y_{i,j} \right\}_{i=1,N}$, for some $j$. These are all
the increments along a row of atoms in the horizontal direction. We hence
freeze the number of particles in the horizontal direction and pass to
the limit in the vertical direction, since we consider the eigenmode of
highest eigenvalue of $\P$ and $\P^\star$. 
\end{remark}

\bigskip

In order to take benefit from (and proceed further
than)~\eqref{eq:youpi3}, we need to solve the 
eigenvalue problem~\eqref{eq:vp2d}. This problem is posed on functions
on $2N$ variables. Solving this problem directly is hence out of reach
numerically. The
approach we suggest consists in further approximating the
problem, arguing somewhat similarly to a mean-field theory. 
Using a sequence of simplifying assumptions (see in
particular~\eqref{eq:prod2d} below), we are going to derive an
integrated form of~\eqref{eq:vp2d} (see~\eqref{eq:g2d} and~\eqref{eq:g*2d}
below). Eventually, this will allow us to recover a 
Markov structure on {\em low} dimensional objects. The invariant measure of
such Markov chain will hence be the solution of an eigenvalue problem,
similarly to~\eqref{eq:vp2d}, but, in contrast to~\eqref{eq:vp2d}, in a
low-dimensional setting, and thus amenable to numerical computations.

\smallskip

Before proceeding along these lines, we make the following
observation. 
We note that any $\mu$ satisfying~\eqref{eq:vp2d} is of the form 
\begin{multline}
\label{eq:mu2d}
\forall (x,y) \in \RR^N \times \RR^N, \quad \mu(x,y) = F(x) F(y) \nu(x), 
\quad \text{with} 
\\
\forall x \in \RR^N, \quad \lambda \, \nu(x) = \int_{\RR^N}
F(z) F(x+z-\tau z) \nu(x+z-\tau z) dz.
\end{multline}
Similarly, any $\mu^\star$ solution to \eqref{eq:vp2d} is of the form
\begin{multline}
\label{eq:mu*2d}
\forall (x,y) \in \RR^N \times \RR^N, \quad 
\mu^\star(x,y) = F(x+\tau y - y) \ \nu^\star(x+\tau y - y), 
\quad \text{with} 
\\ 
\forall x \in \RR^N, \quad 
\lambda \, \nu^\star(x) = \int_{\RR^N} F(z) \ F(x+\tau z -z) \ 
\nu^\star(x+\tau z -z) \ dz.
\end{multline}
In the sequel, we describe a strategy to approximate
$\nu$ and $\nu^\star$, which is based on a direct product
ansatz. The functions $\nu$ and $\nu^\star$ depend on fewer variables
than $\mu$ and $\mu^\star$, and are hence easier to handle.

\subsection{A direct product ansatz}
\label{sec:direct}

Considering that the problems~\eqref{eq:mu2d} and~\eqref{eq:mu*2d} are
(at least asymptotically) symmetric in the variables, we introduce the
ansatz 
\begin{equation}
\label{eq:prod2d}
\text{\textbf{[H3]}} \quad \nu(x) = \prod_{i=1}^N g(x_i), \quad 
\nu^\star(x) = \prod_{i=1}^N g^\star(x_i),
\end{equation}
insert it in~\eqref{eq:mu2d} and~\eqref{eq:mu*2d}, and integrate with
respect to all $x_i$ except one. This gives the following eigenproblems
on $g$ and $g^\star$:
\begin{eqnarray}
\label{eq:g2d}
\lambda g(x_i) \hspace{-3mm}
&=& \hspace{-4mm}
\int_{\RR^2} \hspace{-1mm} f(x_i + z_i - z_{i-1}) f(z_i)
f(z_{i-1}) g(x_i +z_i - z_{i-1}) dz_{i-1} dz_i,
\\
\label{eq:g*2d}
\lambda g^\star(x_i) \hspace{-3mm} 
&=& \hspace{-4mm}
\int_{\RR^2} \hspace{-1mm} f(x_i + z_{i-1} - z_i) f(z_i)
f(z_{i-1}) g^\star(x_i +z_{i-1} - z_i) dz_i dz_{i-1},
\end{eqnarray}
where $\lambda$ is the largest possible eigenvalue. Note
that~\eqref{eq:g2d} and~\eqref{eq:g*2d} are the same equation. Using the 
Krein-Rutman theorem~\cite{schaefer}, it is possible to prove that for
many types of interactions, the corresponding eigenvalue is simple. Hence, 
$g=g^\star$. In addition, observe that~\eqref{eq:g2d} is an eigenvalue
problem on functions of one variable: it can be solved numerically. 

Using~\eqref{eq:mu2d},~\eqref{eq:mu*2d} and~\eqref{eq:prod2d},
we thus have, up to a normalization factor, that, for any 
$(x,y) \in \RR^{2N}$,
\begin{equation}
\label{eq:ansatz_mu}
\mu(x,y) = \prod_{j=1}^N f(x_j) f(y_j) g(x_j), \ \ \mu^\star(x,y) =
\prod_{j=1}^N f(x_j + y_{j+1} - y_j) g(x_j+y_{j+1} - y_j),
\end{equation}
where $g$ is the solution of~\eqref{eq:g2d}. 

\begin{remark}
\label{rem:corr}
Note that the direct product ansatz~\eqref{eq:prod2d} does not imply
that, in the law~\eqref{eq:argh} of $(x_{:,l} , y_{:,l})$, the
increments $(x_{j,l} , y_{j,l})$, $j=1,\ldots,N$, are
independent. Indeed, in view of~\eqref{eq:ansatz_mu}, the numerator 
of~\eqref{eq:argh} reads
\begin{multline*}
\mu(x_{:,l}, y_{:,l}) \ \mu^\star(x_{:,l}, y_{:,l})
= \\
\prod_{j=1}^N f(x_{j,l}) f(y_{j,l}) g(x_{j,l}) 
f(x_{j,l} + y_{j+1,l} - y_{j,l}) g(x_{j,l}+y_{j+1,l} - y_{j,l}),
\end{multline*}
which is {\em not} a direct product of functions depending on
$(x_{j,l},y_{j,l})_{j=1}^N$. 
\end{remark}

We now insert~\eqref{eq:ansatz_mu} in~\eqref{eq:youpi3} and obtain
\begin{multline}
\label{eq:markov-2d}
x^\star = 
\lim_{N\to +\infty} 
Z_N^{-1} \int_{\RR^N \times \RR^N} x_k \ 
\prod_{j=1}^N f(x_j) f(y_j) g(x_j) 
\\
\times f(x_j + y_{j+1} - y_j) g(x_j + y_{j+1} - y_j) \ dx dy,
\end{multline}
where $\dis Z_N = \int_{\RR^N \times \RR^N} \prod_{j=1}^N f(x_j) f(y_j) g(x_j) 
f(x_j + y_{j+1} - y_j) g(x_j + y_{j+1} - y_j) \ dx dy$. 

The Markov structure present in the right hand side
of~\eqref{eq:markov-2d} now allows us to proceed. We introduce the
transfer operator $\Q$ defined by
\begin{multline}
\label{eq:trans2d2}
(\Q\psi)(x_{j+1}, y_{j+1}) = f(x_{j+1}) \ f(y_{j+1}) \int_{\RR^2}
g(x_j) 
\\
\times 
f(x_j + y_{j+1} - y_j) \ g(x_j + y_{j+1} - y_j)
\ \psi(x_j,y_j) \ dx_j \ dy_j,
\end{multline}
its adjoint $\Q^\star$, and the invariant measure equations 
\begin{equation}
\label{eq:vp2d2}
\left\{ \begin{array}{l}
\Q\psi = \eta \psi, 
\\ 
\Q^\star \psi^\star = \eta \psi^\star,
\end{array}\right.\quad \eta = \sup
\text{Spectrum} (\Q) = \sup
\text{Spectrum} (\Q^\star). 
\end{equation}
Using $\Q$ and $\Q^\star$, we recast~\eqref{eq:markov-2d} as
\begin{equation}
\label{eq:youpi4}
x^\star = \lim_{N\to +\infty} 
\frac{\dis
\int_{\RR^2} x_k \ \left( \Q^{k-1} \psi_0 \right)(x_k,y_k)
\ \left( (\Q^\star)^{N-k} \psi_1 \right)(x_k,y_k) \ dx_k \, dy_k
}{\dis
\int_{\RR^2} \left( \Q^{k-1} \psi_0 \right)(x_k,y_k)
\ \left( (\Q^\star)^{N-k} \psi_1 \right)(x_k,y_k) \ dx_k \, dy_k
}
\end{equation}
for some $\psi_0$ and $\psi_1$.
Using again the Krein-Rutman theorem, we obtain that the highest
eigenvalue of $\Q$ is simple and isolated, hence the iterations of $\Q$
converge, up to renormalization, to the projection on $\psi$, the eigenvector
associated to the highest eigenvalue. We hence infer
from~\eqref{eq:youpi4} that 
\begin{equation}
\label{eq:xy*2d}
\begin{array}{rcl}
x^\star &=& 
\frac{\dis
\int_{\RR^2} x_k \, \psi(x_k,y_k) \, \psi^\star(x_k,y_k) \, dx_k \, dy_k}
{\dis \int_{\RR^2} \psi(x_k,y_k) \, \psi^\star(x_k,y_k) \, dx_k \, dy_k},
\\
y^\star &=& 
\frac{\dis
\int_{\RR^2} y_k \, \psi(x_k,y_k) \, \psi^\star(x_k,y_k) \, dx_k \, dy_k}
{\dis \int_{\RR^2} \psi(x_k,y_k) \, \psi^\star(x_k,y_k) \, dx_k \, dy_k}.
\end{array}
\end{equation}

Note that, owing to the specific structure of $\Q$, its eigenvector
$\psi(x,y)$ reads 
\begin{equation}
\label{eq:psi}
\forall (x,y) \in \RR \times \RR, \quad
\psi(x,y) = f(x) \, f(y) \, \overline{\psi}(y),
\end{equation} 
where, for any $y' \in \RR$,
\begin{equation}
\label{eq:toto1}
\eta \overline{\psi}(y') = 
\int_{\RR^2} g(x) \, f(x + y' - y) \, g(x + y' - y) \, f(x) \, 
f(y) \, \overline{\psi}(y) \ dx \ dy.
\end{equation}
Hence, computing the function $\psi$, which depends on {\em two} scalar
variables, amounts to determining $\overline{\psi}$, namely
solving an eigenvalue problem on functions of {\em
  one} variable. In a similar way, the eigenvector $\psi^\star(x,y)$ of
$\Q^\star$ reads 
\begin{equation}
\label{eq:psi_star}
\forall (x,y) \in \RR \times \RR, \quad
\psi^\star(x,y) = g(x) \, \overline{\psi}^\star(x-y),
\end{equation} 
with, for any $x \in \RR$,
\begin{equation}
\label{eq:toto2}
\eta \overline{\psi}^\star(x) = 
\int_{\RR^2} f(x' + y') \, f(y') \, f(x+y') \, 
g(x+y') \, g(x'+y') \, \overline{\psi}^\star(x') \ dx' \ dy'.
\end{equation}
Again, we are left with solving an eigenvalue problem on a function of 
{\em one} variable. 

\subsection{Summary}
\label{sec:algo}

The above approach results in an algorithm for computing an
approximation of $\langle A \rangle_N$. This algorithm may be summarized
as follows:
\begin{enumerate}
\item compute the function $g : \RR \mapsto \RR$ satisfying the
  eigenvalue equation~\eqref{eq:g2d}. Note that $\lambda$ should be the
  largest eigenvalue of the corresponding operator;
\item with $g$, compute the operator $\Q$ given by~\eqref{eq:trans2d2},
  and compute the functions $\psi$ and $\psi^\star$ solution
  to~\eqref{eq:vp2d2}; this amounts in practice to solving the
  eigenvalue problems~\eqref{eq:toto1} and~\eqref{eq:toto2} for
  $\overline{\psi}: \RR \mapsto \RR$ and $\overline{\psi}^\star: \RR
  \mapsto \RR$, where again $\eta$ is the largest eigenvalue; the
  functions $\psi$ and $\psi^\star$ are next computed
  using~\eqref{eq:psi} and~\eqref{eq:psi_star}; 
\item with $\psi$ and $\psi^\star$, compute $x^\star$ and $y^\star$
  given by~\eqref{eq:xy*2d};
\item approximate the limit of $\langle A \rangle_N$ by $A(x^\star+y^\star)$,
  following~\eqref{eq:motivation}.
\end{enumerate}

\begin{remark}
\label{rem:coins2}
As pointed out in Remark~\ref{rem:coins}, our strategy can also handle
observables that depend on $u^{N,N}$, $u^{N,0}$ and $u^{0,N}$. We then
take the approximation
$$
\lim_{N \to \infty}
\langle A \left( u^{N,N},u^{N,0},u^{0,N} \right) \rangle_N \approx
A(x^\star+y^\star,x^\star,y^\star).
$$
\end{remark}

\subsection{The Gaussian case}
\label{sec:gaussian}

In this section, we consider the special case when, in~\eqref{eq:nrj2d},
we choose 
$$
W(x) = x^2, \quad f(x) = \exp(-\beta W(x)) = \exp(-\beta x^2).
$$
Let us show that, in this case, our strategy formally yields the correct
value for $x^\star$ and $y^\star$. To simplify the notation, we assume
henceforth that the inverse temperature is $\beta=1$.

First, since $\mu_N$ is an even function, we infer from~\eqref{eq:x*2d}
that $x^\star = 0$. Let us now follow our strategy, as summarized in
Section~\ref{sec:algo} above. Taking the Fourier transform of the
equation~\eqref{eq:g2d}, we have
$$
\lambda \widehat{g}(\omega) 
=
\widehat{f}(\omega) \widehat{f}(-\omega) \widehat{fg}(\omega).
$$
This is an eigenvalue problem, and we are interested in the case when
$\lambda$ is maximal. In such a case, and under mild assumptions on $f$
(see \cite{bllp} for the details), one easily proves that the corresponding
solution $g$ is unique and never vanishes. In addition, any other
eigenvector must vanish at some point of $\RR$
(see~\cite[Chap. 1, Theorem 1.2]{yihong_du}). Hence, if we find a 
non-vanishing solution of the above problem, it is the unique solution
we are looking for.
Postulating that $g(x) = \exp(-\alpha x^2)$ for some $\alpha$, we deduce
from the above relation that $\alpha$ should satisfy 
$1 = 2 \alpha + 2 \alpha^2$. This equation has two real solutions, and
only one of them is positive. It thus determines $\alpha>0$. This gives
us a positive solution, which, according to the above argument,
corresponds to the largest possible $\lambda$.

We now consider the eigenvalue problem~\eqref{eq:toto1}. Taking the
Fourier transform of this equation, and postulating
that $\overline{\psi}(x) = \exp(-\tau x^2)$ for some $\tau$, we find
that $\tau$ should satisfy $1 + \alpha = 2 \tau + 2 \tau^2$. Again, this
equation has two real solutions, and only one of them is positive. It
thus determines $\tau>0$. 

We finally turn to the eigenvalue problem~\eqref{eq:toto2}, and assume
that its solution reads $\overline{\psi}^\star(x) = \exp(-\gamma x^2)$
for some $\gamma$. A tedious computation then shows that $\gamma$ should
satisfy 
$$
\frac{1}{\gamma} = \frac{1}{1+\alpha} + \frac{1}{1+\tau} > 0.
$$
Having identified the functions $g$, $\overline{\psi}$ and
$\overline{\psi}^\star$, we can now evaluate the right-hand side
of~\eqref{eq:xy*2d}. Since $f$, $g$, $\overline{\psi}$ and
$\overline{\psi}^\star$ are even functions, we note that
$$
\psi(x,y) \psi^\star(x,y) 
=
f(x) f(y) \overline{\psi}(y) \ g(x) \overline{\psi}^\star(x-y)
=
\psi(-x,-y) \psi^\star(-x,-y).
$$
Using the change of variable $(x,y) \mapsto (-x,-y)$ in the right-hand
side of~\eqref{eq:xy*2d}, we deduce that
$$
\frac{\dis
\int_{\RR^2} x \, \psi(x,y) \, \psi^\star(x,y) \, dx \, dy}
{\dis \int_{\RR^2} \psi(x,y) \, \psi^\star(x,y) \, dx \, dy}
=
0.
$$
Consequently, our strategy yields the exact result, namely $x^\star =
0$. In the following section, we numerically consider a more generic
case, for which no analytical solution is available.

\section{Numerical results}
\label{sec:num}

For our numerical simulations, we choose the interaction potential
$$
W(x) = \frac12 (x-1)^4 + \frac12 x^2.
$$
Note that $W$ is not an even function, neither does it satisfy $\dis
\int_\RR x \exp(-\beta W(x)) \, dx = 0$. Hence the difference in height
over a given edge of the membrane lattice does not average to 0.
For a given choice of the inverse temperature $\beta$, we have followed
the strategy summarized in Section~\ref{sec:algo}, that yields $x^\star$
and $y^\star$ defined by~\eqref{eq:xy*2d}. This allows to compute the
reduced model results, that is, the quantity $A(x^\star+y^\star)$, for
any given observable $A$.  

The reference model result is $\langle A \rangle_N$ defined
by~\eqref{eq:aim}. To compute this quantity, we resort to the stochastic 
differential equation
\begin{equation}
\label{eq:sde}
du_t = - \nabla_u E(u_t) \, dt + \sqrt{2 \beta^{-1}} \ dB_t
\quad \text{in} \ \RR^{(N+1)^2-1},
\end{equation}
where $B_t$ is a standard Brownian motion in dimension $(N+1)^2-1$. Of
course, this is a very expensive way of computing $\langle A \rangle_N$,
and we use it only to validate our approach, which is far less
computationally demanding (see Table~\ref{tab:cpu} below, where we compare
computational costs). As recalled in the Introduction, under mild
assumptions on the potential $W$ and the observable $A$, that are here
satisfied, an ergodic theorem holds, and the canonical
average~\eqref{eq:aim} is given by 
$$
\langle A \rangle_N = \lim_{T \to \infty} \frac{1}{T} \int_0^T A
\left(u^{N,N}_t \right) \, dt,
$$
for (almost) all initial conditions $u_{t=0}$. In
practice, the equation~\eqref{eq:sde} is numerically integrated with the
forward Euler scheme (also called the Euler-Maruyama scheme)
\begin{equation}
\label{eq:titi1}
u_{m+1} = u_m - \Delta t \, \nabla_u E(u_m) + \sqrt{2 \Delta t \beta^{-1}} \ G_m
\end{equation}
where $G_m$ is a vector of Gaussian random variables in dimension
$(N+1)^2-1$, and $\Delta t$ is a small time step. In turn, the canonical
average is approximated by
\begin{equation}
\label{eq:titi2}
\langle A \rangle_N \approx \lim_{M \to \infty} \frac1M \sum_{m=1}^M A
\left(u^{N,N}_m \right).
\end{equation}
In practice, we have integrated~\eqref{eq:sde} (using the
scheme~\eqref{eq:titi1}) 
for a large but finite number $M$ of time steps. We thus take the
approximation
\begin{equation}
\label{eq:final_approx}
\langle A \rangle_N \approx \frac1M \sum_{m=1}^M A
\left(u^{N,N}_m \right)
\end{equation}
for large $M$ and small $\Delta t$. The deterministic quantity $\langle
A \rangle_N$ is thus approximated by a random number. To compute error
bars for $\langle A \rangle_N$ (that is, confidence intervals), we have
simulated many independent realizations of the dynamics~\eqref{eq:sde}. 

The reference model results reported here have been obtained
using~\eqref{eq:titi1}-\eqref{eq:titi2} with $\Delta t = 10^{-3}$ and a
number of time steps $M$ large enough such that convergence is reached
in~\eqref{eq:titi2}, up to statistical noise. 

\subsection{Average of observables: convergence with $N$}

\begin{figure}[htbp]
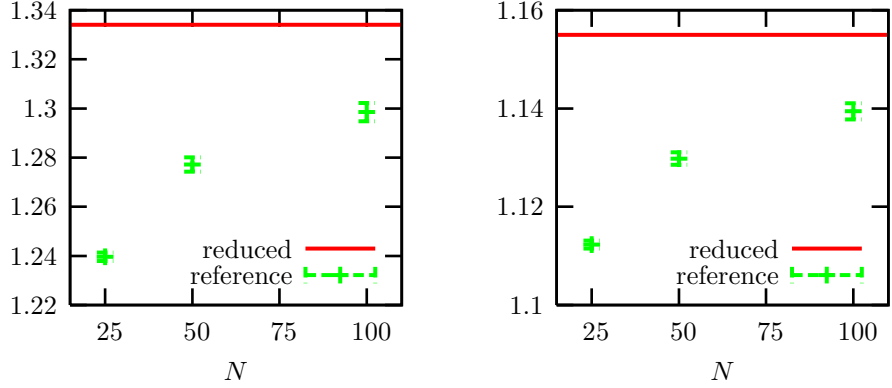

\centerline{
\input{article_N_depend.tex}
\input{article2_N_depend.tex}
}
\caption{Comparison of $\langle A \rangle_N$ (reference result, defined
  by~\eqref{eq:aim}), with $A(x^\star+y^\star)$ (reduced model result), for
  several values of $N$ ($\beta = 1$; left: $A(u) = u^2$; right: $A(u) =
  u$).}
\label{fig:dep1_N}
\end{figure}

\begin{figure}[htbp]
\centerline{
\input{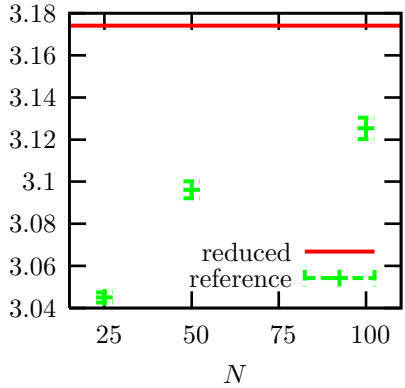}
}
\caption{Comparison of $\langle A \rangle_N$ (reference result defined
  by~\eqref{eq:aim}), with $A(x^\star+y^\star)$ (reduced model result),
  for several values of $N$ ($A(u) = \exp(u)$, $\beta = 1$).}
\label{fig:dep3_N}
\end{figure}

In Figures~\ref{fig:dep1_N}
and~\ref{fig:dep3_N}, 
we compare the reference observable average $\langle A \rangle_N$
with its approximation $A(x^\star+y^\star)$, for several observables $A$, and
for increasing values of $N$ (the temperature is fixed at $\beta^{-1} =
1$). We observe a good agreement
between $A(x^\star+y^\star)$ and $\langle A \rangle_N$, the latter seeming
indeed to converge to the former, when $N \to \infty$: for $N=100$, the
relative difference is of about $1\%$.

\begin{figure}[htbp]
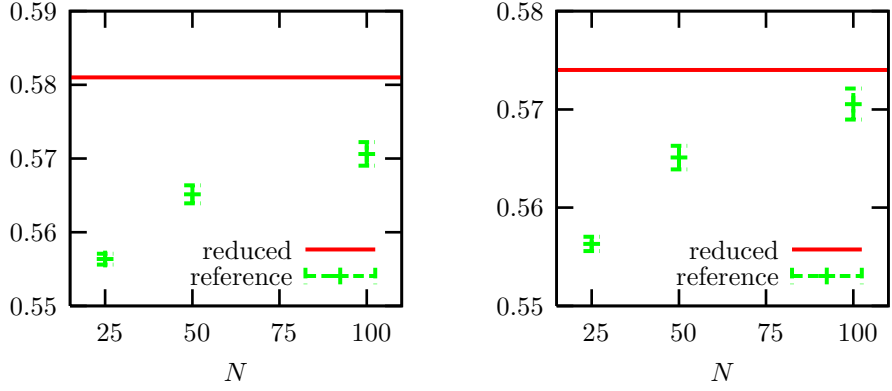

\centerline{
\input{article4_N_depend.tex}
\input{article5_N_depend.tex}
}
\caption{Left: Comparison of $\langle u^{N,0} \rangle_N$ (reference
  result) with $x^\star$ (reduced model result), for several values of
  $N$ ($\beta = 1$). Right: Comparison of $\langle u^{0,N} \rangle_N$
  (reference result) with $y^\star$ (reduced model result).
} 
\label{fig:dep4_N}
\end{figure}

In Figure~\ref{fig:dep4_N}, we compare $\langle u^{N,0} \rangle_N$ with
$x^\star$, in line with Remark~\ref{rem:coins2}. We again observe a good
agreement between the reduced model and the reference model results. On
the same figure,
we compare $\langle u^{0,N} \rangle_N$ with $y^\star$, with similar
conclusions. 

For the sake of completeness, we compare in Table~\ref{tab:cpu} the
costs for computing the reference value $\langle A
\rangle_N$ with those associated to the reduced model approach. We
clearly see that the latter are much smaller than the former. 

\begin{table}[htbp]
\centerline{
\begin{tabular}{| c | c | c | c | c |}
\hline
& $N=25$ & $N=50$ & $N=100$ & Reduced model
\\
\hline
CPU time (days) & 0.2 & 0.7 & 4.6 & 0.025
\\
\hline
\end{tabular}
} 
\caption{First 3 columns: CPU time for computing the reference value $\langle A
\rangle_N$ (for {\em one} choice of observable $A$, at {\em one} given
temperature) using the approximation~\eqref{eq:final_approx} along {\em
  one} single realization of~\eqref{eq:titi1}, with $\Delta t =
10^{-3}$ and $M=10^8$ time steps. 
Last column: CPU time needed by the algorithm of
Section~\ref{sec:algo} (in practice, eigenproblems have been solved on
the interval $I = [-4;5]$, using a spectral basis of 20 functions; integrals
on the interval $I$ have been computed using the 6 Gauss points
quadrature rule, discretizing $I$ into 20 subintervals).
} 
\label{tab:cpu}
\end{table}

\subsection{Average of observables: temperature dependence}

\begin{figure}[htbp]
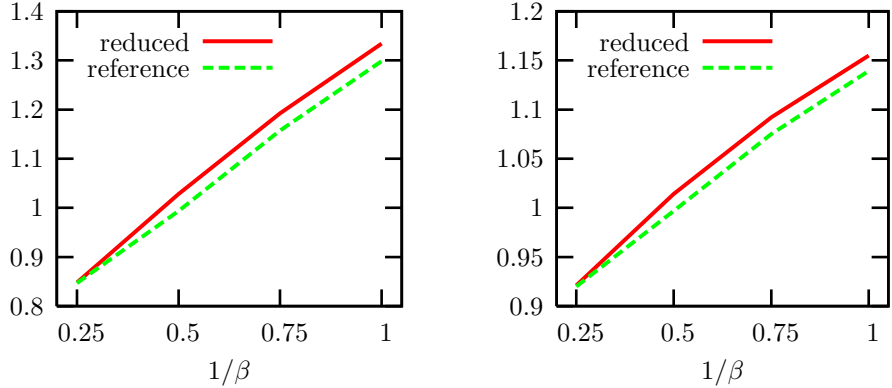

\centerline{
\input{article3_temp_depend.tex}
\input{article1_temp_depend.tex}
}
\caption{Comparison of $\langle A \rangle_N$ (reference result defined
  by~\eqref{eq:aim}), with $A(x^\star+y^\star)$ (reduced model result),
  as a function of the inverse temperature $\beta$ 
  ($N = 100$; left: $A(u) = u^2$; right: $A(u) = u$).} 
\label{fig:dep1_temp}
\end{figure}

\begin{figure}[htbp]
\centerline{
\input{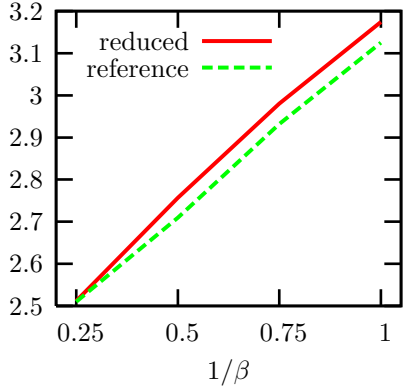}
}
\caption{Comparison of $\langle A \rangle_N$ (reference result defined
  by~\eqref{eq:aim}), with $A(x^\star+y^\star)$ (reduced model result),
  as a function of the inverse temperature $\beta$ ($A(u) = \exp(u)$, $N
  = 100$).}
\label{fig:dep3_temp}
\end{figure}

In Figures~\ref{fig:dep1_temp}
and~\ref{fig:dep3_temp}, 
we compare the reference observable average $\langle A \rangle_N$
with its approximation $A(x^\star+y^\star)$, for several observables $A$, and
for different temperatures (the reference system is run with $N=100$
particles in each direction). We observe a good agreement
between $A(x^\star+y^\star)$ and $\langle A \rangle_N$, for all the temperatures
considered.

\begin{figure}[htbp]
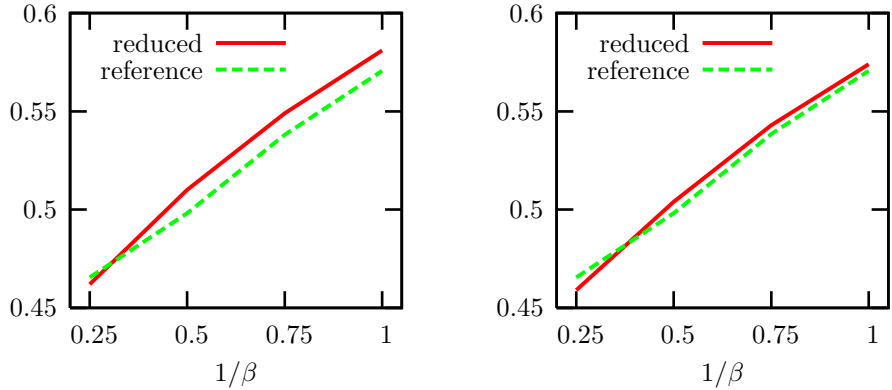

\centerline{
\input{article2_temp_depend.tex}
\input{article5_temp_depend.tex}
}
\caption{Left: Comparison of $\langle u^{N,0} \rangle_N$ (reference
  result) with $x^\star$ (reduced model result), as a
  function of the inverse temperature $\beta$ ($N = 100$).
Right: Comparison of $\langle u^{0,N} \rangle_N$ (reference
  result) with $y^\star$ (reduced model result).
}  
\label{fig:dep4_temp}
\end{figure}

On Figure~\ref{fig:dep4_temp}, we compare $\langle u^{N,0} \rangle_N$
(respectively $\langle u^{0,N} \rangle_N$) with $x^\star$ (respectively
$y^\star$). We again observe a good agreement between the reduced model
and the reference model results, for all the temperatures we have
tested.

\subsection{Correlations}
\label{sec:corr}


In this section, we compare the correlations of successive
increments in the membrane. More precisely, we compute, according to the
reference model and the reduced model, the average of products of the
type $x_{k,l} x_{k+p,l}$ for some $p \geq 0$. Recall that
$$
x_{k,l} = \frac{u^{k,l} - u^{k-1,l}}h
$$
is the (rescaled) difference in height between atoms at sites $(k-1,l)$
and $(k,l)$. This kind of tests are more demanding than the ones
considered above.

The reference model computations have been performed on a system of
$(N+1)^2$ atoms. The correlation can be defined from all the
increments except those at the boundaries, or just the increments at the
center of the system. Both yield very close results (the difference is
smaller than 0.2 \%). In addition,
those reference computations have been computed by averaging several
independent realizations. The number of realizations we have considered
is large enough so that the statistical noise is smaller (in relative
error) than 0.2 \%. Results given by the reference computations are
gathered in the second column of Table~\ref{tab:corr}. 

\medskip

The reduced model values have been computed by starting from the exact 
values, e.g.
$$
\langle x_{k,l}^2 \rangle = 
Z_N^{-1} \int x^2_{k,l} \ 
\mu_N \left(\left\{x_{i,j}\right\}, \left\{y_{i,j}\right\} \right),
$$
and following the steps described in
Section~\ref{sec:theo}. This yields the quantity
\begin{equation}
\label{eq:pas_grosse}
\langle x_{k,l}^2 \rangle_{\rm red} = 
\frac{\displaystyle 
\int_{\RR^2} x^2_k \, \psi(x_k,y_k) \, \psi^\star(x_k,y_k) \, dx_k \, dy_k}
{\displaystyle \int_{\RR^2} \psi(x_k,y_k) \, \psi^\star(x_k,y_k) \, dx_k \,
  dy_k}
\end{equation}
as an approximation of $\langle x_{k,l}^2 \rangle$. In turn, the
quantity $\langle x_{k,l} \ x_{k+1,l} \rangle$ is approximated by
\begin{multline}
\label{eq:grosse}
\langle x_{k,l} \ x_{k+1,l} \rangle_{\rm red} = 
\\
\frac{\displaystyle 
\int_{\RR^4} x_k x_{k+1} \, \psi(x_k,y_k) 
\, c(x_k,y_k,x_{k+1},y_{k+1})
\, \psi^\star(x_{k+1},y_{k+1}) \,
dx_k \, dy_k \, dx_{k+1} \, dy_{k+1}}
{\displaystyle \int_{\RR^4} \psi(x_k,y_k) \, c(x_k,y_k,x_{k+1},y_{k+1})
\, \psi^\star(x_{k+1},y_{k+1}) \,
dx_k \, dy_k \, dx_{k+1} \, dy_{k+1}},
\end{multline}
with 
\begin{multline*}
c(x_k,y_k,x_{k+1},y_{k+1}) = 
\\
g(x_k) \, f(x_{k+1}) \, f(y_{k+1}) 
\, g(x_k + y_{k+1} - y_k) \, f(x_k + y_{k+1} - y_k).
\end{multline*}
Note that formulas \eqref{eq:pas_grosse} and \eqref{eq:grosse} are
consistent with each other. Indeed, if $x_{k+1,l}$ is replaced by
$x_{k,l}$ in~\eqref{eq:grosse}, we have
\begin{eqnarray*}
&& \langle x_{k,l} \ x_{k,l} \rangle_{\rm red} 
\\
&=& 
\frac{\displaystyle 
\int_{\RR^4} x_k^2 \, \psi(x_k,y_k) 
\, c(x_k,y_k,x_{k+1},y_{k+1})
\, \psi^\star(x_{k+1},y_{k+1}) \,
dx_k \, dy_k \, dx_{k+1} \, dy_{k+1}}
{\displaystyle \int_{\RR^4} \psi(x_k,y_k) \, c(x_k,y_k,x_{k+1},y_{k+1})
\, \psi^\star(x_{k+1},y_{k+1}) \,
dx_k \, dy_k \, dx_{k+1} \, dy_{k+1}}
\\
&=& 
\frac{\displaystyle 
\int_{\RR^2} x_k^2 \, \psi(x_k,y_k) 
\, (\Q^\star \psi^\star)(x_k,y_k) \, dx_k \, dy_k}
{\displaystyle \int_{\RR^2} \psi(x_k,y_k) \, 
(\Q^\star \psi^\star)(x_k,y_k) \, dx_k \, dy_k}
\\
&=& 
\frac{\displaystyle 
\int_{\RR^2} x_k^2 \, \psi(x_k,y_k) 
\, \psi^\star(x_k,y_k) \, dx_k \, dy_k}
{\displaystyle \int_{\RR^2} \psi(x_k,y_k) \, 
\psi^\star(x_k,y_k) \, dx_k \, dy_k}
\end{eqnarray*}
and we recover~\eqref{eq:pas_grosse}. We have used in the above
derivation the fact that
\begin{multline*}
\int_{\RR^2} c(x_k,y_k,x_{k+1},y_{k+1}) 
\, \psi^\star(x_{k+1},y_{k+1}) \, dx_{k+1} \, dy_{k+1}
=\\
\Q^\star \psi^\star(x_k,y_k) = \eta \psi^\star(x_k,y_k).
\end{multline*}
  
Integrals in dimension 2 and 4 need to be evaluated to
compute~\eqref{eq:grosse} (and similarly, integrals in dimension 6 need
to be evaluated 
to compute $\langle x_{k,l} \ x_{k+2,l}  \rangle_{\rm red}$). These
integrals have been computed almost exactly using numerical
quadratures. We gather the obtained results in the third column of
Table~\ref{tab:corr}.

\begin{table}[htbp]
\centerline{
\begin{tabular}{| c | c | c | c |}
\hline
& Reference model & Reduced model & Relative difference
\\
\hline
$\langle x_{k,l}^2 \rangle$ & 0.50668  & 0.52808 & $4.05 \times 10^{-2}$
\\
\hline
$\langle y_{k,l}^2 \rangle$ & 0.50684 & 0.51217 & $1.04 \times 10^{-2}$
\\
\hline
$\langle x_{k,l} \ x_{k+1,l} \rangle$ & 0.28097 & 0.31697 & 0.114
\\
\hline
$\langle x_{k,l} \ x_{k+2,l} \rangle$ & 0.31215 & 0.33089 & $5.66 \times 10^{-2}$
\\
\hline
$\langle x_{k,l} \rangle$ & 0.57310 & 0.581 & $1.36 \times 10^{-2}$
\\
\hline
\end{tabular}
} 
\caption{Comparison of the correlations of successive increments
  according to the reduced model with the reference model values (inverse
  temperature $\beta = 1$, $N = 100$, $1 \ll k,l \ll N$). The reference
  values are expectations (e.g. of $x_{k,l}^2$) with respect to the Gibbs
measure $Z_N^{-1} \overline{\mu}_N$ defined by~\eqref{eq:gibbs}.} 
\label{tab:corr}
\end{table}
%

We observe that the difference between the reference result and the
reduced model results is of the order of 5\% for all values except one. We consider that such an accuracy
is good, given the number of assumptions on which the numerical strategy
leading to the reduced model values is based.

\section*{Acknowledgements}

The present contribution is related to a lecture given by FL at 
a workshop at Leuven on ``Multiscale simulation of heterogeneous
materials'' (January 12-14, 2011). FL would like to thank the organizers
of the workshop for their kind invitation.
This work is supported in part by the INRIA, under the grant ``Action
de Recherche Collaborative'' HYBRID, and by the Agence Nationale de la
Recherche, under grant ANR-09-BLAN-0216-01 (MEGAS).

\end{document}